\documentclass[%
 reprint,
 superscriptaddress,
 amsmath,amssymb,
 aps,
 pra,
]{revtex4-1}

\usepackage{graphicx}
\usepackage{dcolumn}
\usepackage{bm}
\usepackage{hyperref}
\usepackage{url}
\usepackage{comment}

\newcommand{\eref}[1]{Eq.~(\ref{#1})}

\newcommand{\fref}[1]{Fig.~\ref{#1}}
\newcommand{\Fref}[1]{Figure~\ref{#1}}

\newcommand{\tref}[1]{table~\ref{#1}}

\begin{document}
\newcommand{\ManuscriptTitle}{
    A Simple Data-Driven Level Finding Method of Quantum Many-Body Systems 
    \\ based on Statistical Outlier Detection
}

\title{\ManuscriptTitle}


\author{Kazuaki Hongu}
\author{Keisuke Fujii}
\email{fujii.keisuke.2z@kyoto-u.ac.jp}
\affiliation{%
Department of Mechanical Engineering and Science,
Graduate School of Engineering, Kyoto University
Kyoto 615-8540, Japan}

\date{\today}

\begin{abstract}    
We report a simple and pure data-driven method to find new energy levels of quantum many-body systems only from observed line wavelengths. 
In our method, all the possible combinations are computed from known energy levels and wavelengths of unidentified lines.
As each excited state exhibits many transition lines to different lower levels, the true levels should be reconstructed coincidentally from many level-line combinations, while the wrong combinations distribute randomly.
Such a coincidence can be easily detected statistically.
We demonstrate this statistical method by finding new levels for various atomic and nuclear systems from unidentified line lists available online.
\end{abstract}

\maketitle

Energy levels are one of the most important quantities of quantum systems.
From Ritz~\cite{ritz_new_1908}, the energy structures of various quantum systems, such as atoms and nuclei, have been determined from the emission observation, the wavelength of which corresponds to the energy interval of the transition.
As more than thousands of lines are observed from many-electron atoms and heavy nuclei, it is not straightforward to find the correct combinations of these lines and levels and reconstruct their energy structure. 

For atomics systems, highly accurate \textit{ab initio} simulation codes have been often used to help the identification~\cite{cowan_theory_1981, Bar-Shalom2001, Gu2008, kramida_cowan_2019}.
By fine-tuning the simulation parameters and comparing with the observed emission wavelengths and intensities, the energy levels of many systems have been determined~\cite{lewis_spectral_1987,bekker_detection_2019,arab_observation_2019}.
As this process requires a lot of try-and-errors of experienced scientists, some methods have been proposed to speed up the identification~\cite{azarov_formal_1991,azarov_formal_1993} and some have been widely used for the level identification of many-electron atoms~\cite{arab_observation_2019}.
Other experimental quantities have been also used, such as intensities, isotope shifts, and Zeeman-effect, to help the identification.

For nuclear systems, the coincidence technique has been used to identify their energy structure~\cite{melissinos_experiments_2003}.
From the energies of two gamma photons observed from an excited nucleus in a very short interval, we find that these transitions belong to a single decay path.
A data analysis technique to speed up the level identification has been also proposed~\cite{demand_transition-centric_2013}.
Still, the level identification is time consuming particularly for complex nuclei and highly excited states. 

For both fields, methods independent from accurate simulations and sophisticated experimental techniques would complement the existing level-identification methods.
Here, we report a very simple method that relies only on a list of the observed wavelengths.
Rather than seeking the optimal combination, we computed all the level candidates from the known lower levels and observed wavelengths.
The problem we consider here is 
to find new upper-level energies $\mathbf{x} = \{E_{m_1} + \lambda_{n_1}, \cdots\}$, 
from known lower-level energies $\mathbf{E} = \{E_1, E_2, \cdots, E_M\}$ 
and experimentally observed energy intervals $\mathbf{L} = \{\lambda_1, \cdots, \lambda_N\}$,
where $\{(m_1, n_1), \cdots\}$ indicate the correct identification of lines and levels.
In the proposed method, we simply compute all the level candidates by
$\mathbf{y} = \{y_1, \cdots, y_{NM}\} = \mathbf{E} \cdot \mathbf{1}^\top + \mathbf{1} \cdot \mathbf{L}^\top = \{E_m + \lambda_n | m\in [1,M], n\in [1,N]\}$.
As most of the entries of $\mathbf{y}$ are constructed from wrong combinations, they distribute randomly.
On the other hand, some combinations are correct and as typically there are many transitions from a single upper level, the true combinations for these transitions should coincidently point the energy of the true upper level.
Such a coincidence can be distinguished from the randomly distributed candidates based on widely-used outlier-detection methods.

\begin{figure*}[ht]
    \includegraphics[width=18cm]{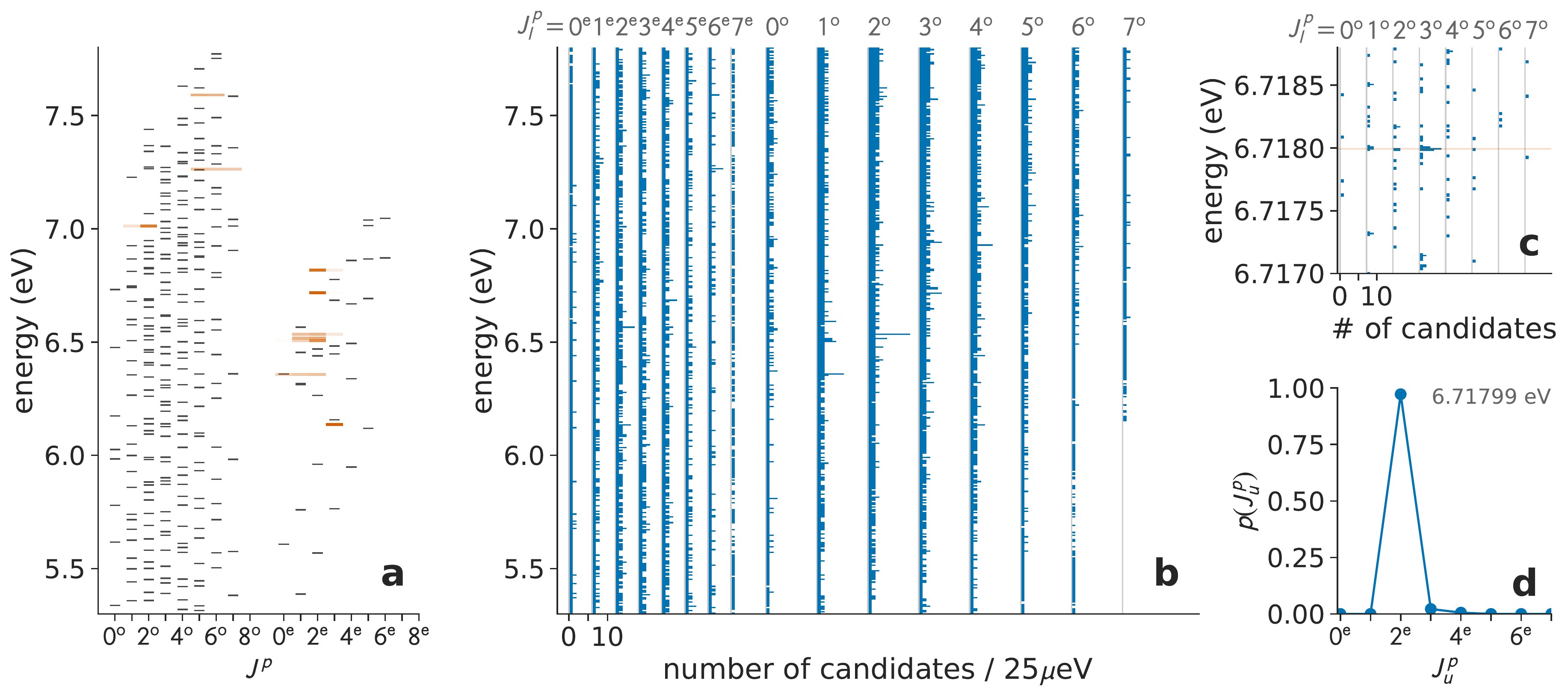}
    \caption{%
    (a) The energy level diagram of neutral tungsten atom.
    The gray bars indicate the known levels included in the ASD~\cite{NIST_ASD}.
    The red bars are the new energy levels found in this work, where the uncertainty of the $J$ values are indicated by the transparency of the markers.
    (b)
    Histograms of the energy level candidates constructed from the known levels and unidentified transition energies.
    The energy bin size for the histogram is 25 $\mu$eV.
    The $J^p$ values of the lower levels are indicated at the top of each trace.
    Several peaks can be seen, indicating the new energy levels.
    (c)
    An expanded view of (b) around $E \approx 6.51799$ eV.
    The horizontal red line indicates the detected new energy level.
    (d)
    The probability distribution of $J_u^p$ values for the new level at $E =$ 6.51799 eV.
    }
    \label{fig:tungsten}
\end{figure*}

As a demonstration, we consider to find new energy levels of neutral tungsten atoms from a public database.
We use the data in the Atomic Spectra Databse (ASD) maintained by National Institute of Standards and Technology~\cite{NIST_ASD}.
In the ASD, 465 energy levels are already identified for the neutral tungsten~\cite{NIST_ASD, kramida_compilation_2006}.
Not only the values of the energy, but also the total angular quantum numbers $J$ and the parities $p$ of these states, both of which are good quantum numbers even with heavy wavefunction mixing, are stored.
The gray horizontal bars in \fref{fig:tungsten}~(a) show these energy levels of the neutral tungsten from the ASD.

The ASD also contains 1197 unidentified lines for the neutral tungsten, ranging from 200.1 nm to 1010.7 nm~\cite{NIST_ASD, laun_first_1968}.
From these known energy levels and unidentified emission line wavelengths, we compute energy level candidates.
Figures~\ref{fig:tungsten}~(b, c) show the number of the level candidates as a function of the excited energy, i.e., the histogram of $\mathbf{y}$. 
These histograms are construted from all the lower levels having $J_l^{p}$, which is the pair of $(J, p)$ of the lower levels.
From the experimental accuracy of the wavelengths, we determine the size of the energy bin $\delta E$ to be $25\,\mu$eV.
Several peaks are apparent in the figure, along with a finite noise floor coming from wrong combinations.

As the energy values from the wrong combinations may distribute randomly, their probability density is approximated by the Poisson distribution, i.e., the probability to have $r_i$ candidates in the $i$-th bin is represented by
\begin{align}
    p_\mathrm{poi}(r_i) = \frac{\rho^{r_i} e^{-\rho}}{r_i!},
    \label{eq:poisson}
\end{align}
where $\rho$ is the average of $r_i$.
We estimate $\rho$ by taking a moving average of $r$ in a certain energy interval.
The estimation uncertainty of $\rho$ makes the predictive distribution of $r$ the negative binomial distribution,
\begin{align}
    p_\mathrm{NB}(r) = \frac{(k + r - 1)!}{k! (r-1)!} 
    \left(1-\frac{1}{K + 1}\right)^{r} \left(\frac{1}{K + 1}\right)^k,
    \label{eq:nbinom}
\end{align}
where $K$ is the number of bins used to estimate $\rho$ while $k$ is the total number of candidates fallen into these bins.
The details of this predictive distribution can be found in the Appendix.

Based on \eref{eq:nbinom}, outliers in the histogram, which indicate the true energy levels, can be detected by standard outlier detection techniques.
In this work for the sake of the simplicity, we detect peaks based on a threshold value.
The $i$-th bin is determined as an outlier if $r_i > \underline{r}$, where $\underline{r}$ is the threshold satisfying
\begin{align}
    \label{eq:threshold}
    \int_{\underline{r}}^\infty p_\mathrm{NB}(x) dx = q \frac{\delta E}{E^\mathrm{max} N_J N_p},
\end{align}
where $E^\mathrm{max}$ is the maximum energy range to be considered, $N_J$ is the number of possible $J$ values, and $N_p = 2$ is the number of possible parities. 
$q$ is the inverse safety factor. 
Essentially, \eref{eq:threshold} indicates that the probability to detect a single false peak for this system is set to be $q$.
For the neutral tungsten, we use $E^\mathrm{max} = 10$ eV and $q = 1/30$.
The list of the new energy levels detected by this method is shown in \tref{tab:W}.
This table also shows the number of lines pointing each level $r$ as well as the logarithmic value of the survival function
\begin{align}
-\ln s = -\ln \int_r^\infty p_\mathrm{NB}(x) dx,
\end{align}
which can be compared with \eref{eq:nbinom}.

The selection rule of the electric dipole transition helps us to determine the parities and total angular quantum numbers of newly detected levels $J_u^p$.
For example, we found a peak at $E$ = 6.71799 eV in the histogram constructed from the lower level with $J_l^p$ = $3^\mathrm{o}$ (\fref{fig:tungsten}~(c)).
Smaller peaks at $J_l^p$ = $1^\mathrm{o}$ and $2^\mathrm{o}$ suggest $J_u^p = 2^\mathrm{e}$ for this new level.

We again use \eref{eq:nbinom} to infer the value of $J_u^p$.
Let us assume that there is an energy level at $E_i$ with $J_u^p$.
With the small enough $\delta E$, there should be no other energy levels in the interval $[E_i, E_i + \delta E)$.
Therefore, the candidates constructed from the lower levels with $J_l^p \in \overline{S(J_l^p)}$ should follow \eref{eq:nbinom}, where 
$S(J_l^p)$ indicates the set of the allowed $J$s and parities by electric dipole transitions, i.e., $S(J_l^p) = \{(J_u - 1)^{-p_u}, J_u^{-p_u}, (J_u + 1)^{-p_u}\}$ if $J_u \neq 0$ and $S(J_l^p) = \{(J_u + 1)^{-p_u}\}$ otherwise.
$\overline{S(J_l^p)}$ is the comprementary set of $S(J_l^p)$.
The probability that the upper level has $J_u^p$ can be formulated as
\begin{align}
    p(J_u^{p}) = \frac{
        p_\mathrm{NB}(r\;_{\overline{S(J_u^p)}})
    }{
        \sum_{J, p} p_\mathrm{NB}(r\;_{\overline{S(J^p)}})
    }.
    \label{eq:probabilityJ}
\end{align}
Here, $r\;_{\overline{S(J_u^p)}}$ is the number of level candidates at a certain energy bin constructed from the lower levels with $J_l^p \in \overline{S(J_u^p)}$.
\Fref{fig:tungsten}~(d) shows the probability distribution of the $J_u^p$ for the new energy level at $E$ = 6.71799 eV, computed from \eref{eq:probabilityJ}.
As expected, we obtain the higher probability of $2^\mathrm{e}$ for this level but $3^\mathrm{e}$ is still slightly probable.
In \fref{fig:tungsten}~(a), we show all the new energy levels detected in this work by red markers, with the uncertainty of $J^p$ indicated by the transparency of the markers.
The inferred $J^p$ for all the detected levels are shown in \tref{tab:W}.

\begin{table}
    \caption{\label{tab:W}New energy levels of W found in this work.}
    \begin{ruledtabular}
    \begin{tabular}{c c c c c}  
      energy (eV) & $J^p$  $_\mathrm{(prob.)}$ & \# of lines & $-\ln s$ \\ [0.5ex] 
      \hline
      6.1356 & $3^\mathrm{e}$  $_{(1.00)}$ & 8 & 22.1\\
      6.3577 & $2^\mathrm{e}$  $_{(0.37)}$, $1^\mathrm{e}$  $_{(0.34)}$, $0^\mathrm{e}$  $_{(0.28)}$ & 7 & 18.5\\
      6.5054 & $2^\mathrm{e}$  $_{(0.70)}$, $1^\mathrm{e}$  $_{(0.28)}$ & 8 & 19.7\\
      6.5180 & $2^\mathrm{e}$  $_{(0.54)}$, $1^\mathrm{e}$  $_{(0.46)}$ & 9 & 22.3\\
      6.5341 & $2^\mathrm{e}$  $_{(0.45)}$, $1^\mathrm{e}$  $_{(0.40)}$, $3^\mathrm{e}$  $_{(0.15)}$ & 12 & 31.7\\
      6.7180 & $2^\mathrm{e}$  $_{(0.97)}$ & 10 & 23.8\\
      6.8169 & $2^\mathrm{e}$  $_{(0.89)}$, $3^\mathrm{e}$  $_{(0.11)}$ & 11 & 26.5\\
      7.0115 & $2^\mathrm{o}$  $_{(0.81)}$, $1^\mathrm{o}$  $_{(0.18)}$ & 6 & 20.2\\
      7.2641 & $5^\mathrm{o}$  $_{(0.35)}$, $6^\mathrm{o}$  $_{(0.33)}$, $7^\mathrm{o}$  $_{(0.32)}$ & 4 & 14.6\\
      7.5907 & $5^\mathrm{o}$  $_{(0.51)}$, $6^\mathrm{o}$  $_{(0.47)}$ & 5 & 18.3\\
    \end{tabular}
    \end{ruledtabular}
    \end{table}
    
\begin{figure}[h]
    \includegraphics[width=8cm]{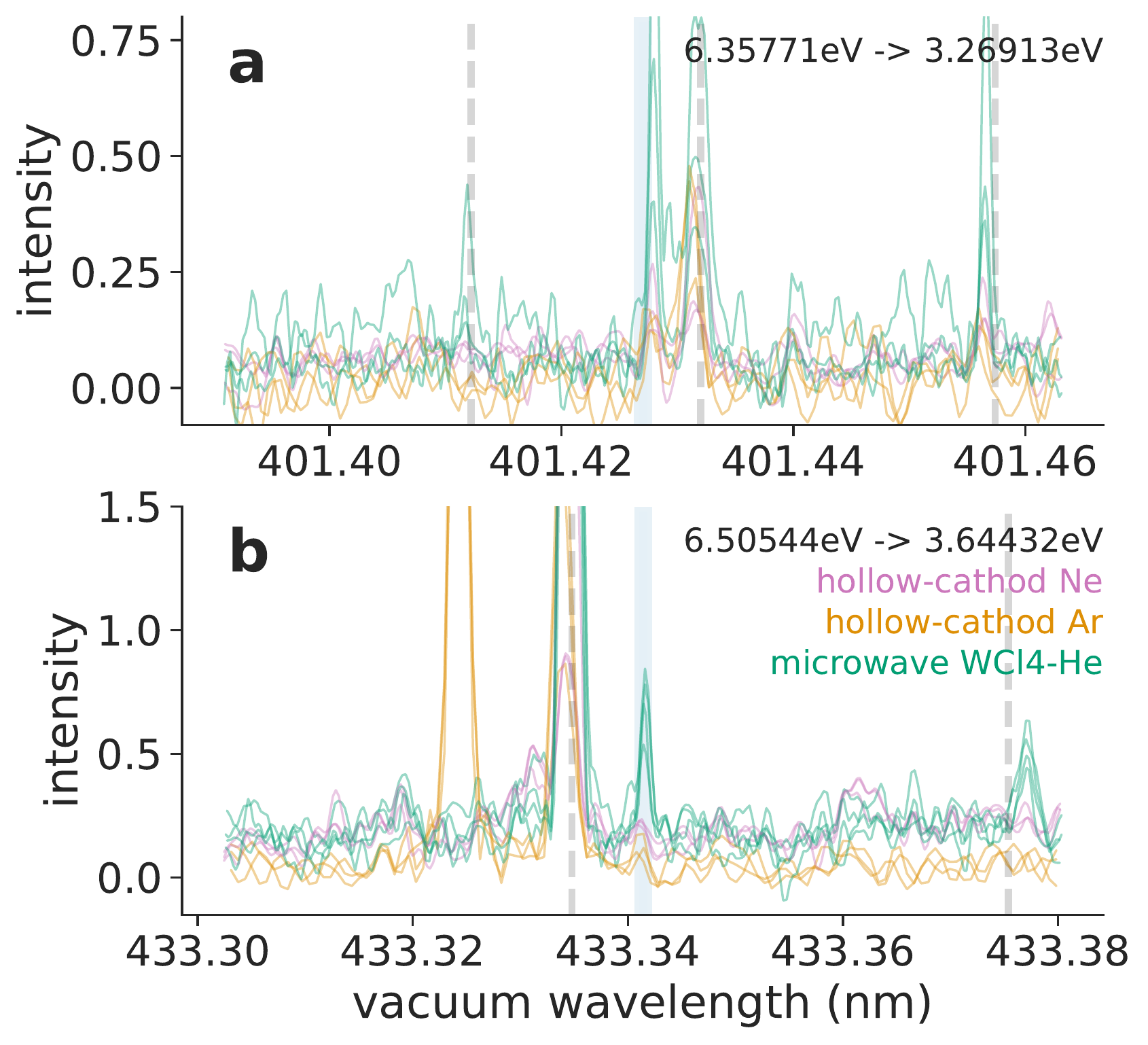}
    \caption{%
        Emission spectra observed at NSO~\cite{NSO} for (pink) hollow-cathode discharges of W with Ne as the buffer gas, 
        (yellow) those with Ar as the buffer gas, and (green) microwave plasma of He and WCl$_4$ mixture.
        Vertical dashed lines indicate known emission lines of W.
        The bold vertical lines indicate possible emission line wavelengths from the newly found levels at (a) 6.35771 eV and (b) 6.50544 eV.
        The width of the shaded region indicates the wavelength uncertainty corresponding to $\delta E$. 
    }
    \label{fig:spectra}
\end{figure}

As an independent validation of this result, we search new emission lines from these newly-found levels in high-resolution spectra stored in the historical archive of the National Solar Observatory (NSO)~\cite{NSO}.
Several types of plasmas containing tungsten atoms, such as hollow-cathode discharges with the buffer gas of neon and argon, as well as the microwave discharge with $\mathrm{W Cl_4}$, have been observed with a high-resolution Fourier transform spectrometer at the NSO.
Figures~\ref{fig:spectra} show a few examples of such spectra, with the discharge types indicated by the color of the traces.
The shaded region in each figure shows the predicted emission line wavelength from the newly detected levels, where its width indicates the uncertainty corresponding to $\delta E$.
The vertical dashed lines show known tungsten emission line wavelengths.
Peaks are apparent at the predicted wavelength for all the types of discharges, although the intensities are weaker than the known lines.
The existence of these lines suggests the validity of the newly found levels and our method proposed here.

As this method is very simple and no \textit{ab initio} computations are required, we can conduct the same analysis for other atoms very quickly.
We repeat the above procedure for neutral silicon, scandium, iron, renium, and osmium, also from the ASD.
Several new levels are found.
The results are shown in \tref{tab:atom}.

\begingroup
\squeezetable
\begin{table}
\caption{\label{tab:atom} New energy levels found in this work. For all the atoms, 
$E^\mathrm{max}$ = 10 eV are used.
$\delta E = 10 \,\mu$eV for Fe and $25 \mu$eV for other atoms.
}
\begin{ruledtabular}
\begin{tabular}{c c c c c c}  
  atom & energy (eV) & $J^p$  $_\mathrm{(prob.)}$ & \# of lines & $-\ln s$ \\ [0.5ex] 
  \hline
Si & 7.8585 & $2^\mathrm{e}$  $_{(0.50)}$, $1^\mathrm{e}$  $_{(0.49)}$ & 3 & 18.1\\
 & 8.7375 & $3^\mathrm{o}$  $_{(0.34)}$, $4^\mathrm{o}$  $_{(0.33)}$, $5^\mathrm{o}$  $_{(0.33)}$ & 3 & 15.7\\
 & 8.8679 & $3^\mathrm{o}$  $_{(0.49)}$, $4^\mathrm{o}$  $_{(0.48)}$ & 3 & 16.7\\
 & 8.9891 & $3^\mathrm{o}$  $_{(0.50)}$, $4^\mathrm{o}$  $_{(0.49)}$ & 4 & 18.1\\
 & 9.4394 & $3^\mathrm{o}$  $_{(0.50)}$, $4^\mathrm{o}$  $_{(0.49)}$ & 4 & 18.2\\
\hline
Sc & 4.3690 & $\frac{3}{2}^\mathrm{e}$  $_{(0.98)}$ & 3 & 19.2\\
 & 4.7566 & $\frac{7}{2}^\mathrm{o}$  $_{(0.50)}$, $\frac{9}{2}^\mathrm{o}$  $_{(0.49)}$ & 3 & 19.0\\
 & 6.0203 & $\frac{7}{2}^\mathrm{o}$  $_{(0.48)}$, $\frac{9}{2}^\mathrm{o}$  $_{(0.46)}$ & 4 & 17.1\\
\hline
Fe & 5.8629 & $4^\mathrm{e}$  $_{(0.99)}$ & 5 & 31.4\\
\hline
Gd & 4.0517 & $2^\mathrm{o}$  $_{(0.34)}$, $1^\mathrm{o}$  $_{(0.33)}$, $0^\mathrm{o}$  $_{(0.32)}$ & 3 & 13.5\\
\hline
Hf & 5.7896 & $5^\mathrm{e}$  $_{(0.73)}$, $4^\mathrm{e}$  $_{(0.25)}$ & 6 & 23.6\\
 & 5.8706 & $4^\mathrm{e}$  $_{(0.47)}$, $5^\mathrm{e}$  $_{(0.42)}$, $3^\mathrm{e}$  $_{(0.10)}$ & 5 & 15.4\\
 & 6.0524 & $3^\mathrm{o}$  $_{(0.97)}$ & 7 & 21.2\\
 & 6.2479 & $4^\mathrm{o}$  $_{(0.79)}$, $3^\mathrm{o}$  $_{(0.15)}$ & 6 & 19.5\\
 & 6.2937 & $4^\mathrm{o}$  $_{(0.47)}$, $5^\mathrm{o}$  $_{(0.43)}$, $3^\mathrm{o}$  $_{(0.10)}$ & 6 & 19.1\\
 & 6.3308 & $4^\mathrm{o}$  $_{(0.52)}$, $5^\mathrm{o}$  $_{(0.47)}$ & 6 & 18.8\\
 & 6.5040 & $3^\mathrm{o}$  $_{(0.53)}$, $4^\mathrm{o}$  $_{(0.46)}$ & 7 & 17.6\\
\hline
Re & 6.7303 & $\frac{7}{2}^\mathrm{e}$  $_{(0.50)}$, $\frac{9}{2}^\mathrm{e}$  $_{(0.49)}$ & 4 & 19.5\\
 & 7.6139 & $\frac{3}{2}^\mathrm{e}$  $_{(0.93)}$ & 5 & 19.1\\
\hline
Os & 5.9089 & $3^\mathrm{e}$  $_{(0.94)}$ & 4 & 19.7\\
 & 6.6527 & $5^\mathrm{e}$  $_{(0.90)}$ & 4 & 19.9\\
 & 7.2361 & $4^\mathrm{e}$  $_{(0.92)}$ & 5 & 19.2\\
 & 4.8857 & $5^\mathrm{o}$  $_{(0.94)}$ & 5 & 26.1\\
 & 5.4383 & $3^\mathrm{o}$  $_{(0.94)}$ & 6 & 24.5\\
\end{tabular}
\end{ruledtabular}
\end{table}
\endgroup

This method is not limited to atomic systems but also applicable to other quantum systems such as nuclei.
We downloaded known excited levels and unidentified $\gamma$ line energies of several heavy nuclei from the ENSDF database by the National Nuclear Data Center~\cite{ENSDF}.
Based on the exactly the same analysis, we found new energy levels for several heavy nuclei.
The results are shown in \tref{tab:nucleus}.
15 new energy levels in total are found with this method.

\begingroup
\squeezetable
\begin{table}
\caption{\label{tab:nucleus} New energy levels of nuclei found in this work. For all the nuclei, 
$E^\mathrm{max}$ = 10 MeV and $\delta E$ = 0.1 keV are used.
}
\begin{ruledtabular}
\begin{tabular}{c c c c c c}  
  nucleus & energy (keV) & $J^p$  $_\mathrm{(prob.)}$ & \# of lines & $-\ln s$ \\ [0.5ex] 
  \hline
$^{28}\mathrm{Al}$ & 6967.6 & $2^\mathrm{e}$  $_{(0.98)}$ & 3 & 17.5\\
\hline
$^{36}\mathrm{Cl}$ & 3207.0 & $2^\mathrm{e}$  $_{(0.98)}$ & 3 & 20.9\\
\hline
$^{60}\mathrm{Ni}$ & 9729.3 & $7^\mathrm{o}$  $_{(0.35)}$, $8^\mathrm{o}$  $_{(0.33)}$, $9^\mathrm{o}$  $_{(0.32)}$ & 3 & 20.4\\
\hline
$^{66}\mathrm{Cu}$ & 3695.3 & $2^\mathrm{e}$  $_{(0.50)}$, $1^\mathrm{e}$  $_{(0.50)}$ & 3 & 16.9\\
\hline
$^{74}\mathrm{Ge}$ & 9864.7 & $4^\mathrm{o}$  $_{(0.88)}$ & 3 & 17.9\\
\hline
$^{82}\mathrm{Br}$ & 8946.4 & $2^\mathrm{e}$  $_{(0.44)}$, $4^\mathrm{e}$  $_{(0.26)}$, $3^\mathrm{e}$  $_{(0.26)}$ & 3 & 6.0\\
\hline
$^{87}\mathrm{Kr}$ & 5251.8 & $\frac{5}{2}^\mathrm{o}$  $_{(0.33)}$, $\frac{3}{2}^\mathrm{o}$  $_{(0.33)}$, $\frac{1}{2}^\mathrm{o}$  $_{(0.33)}$ & 3 & 18.0\\
\hline
$^{90}\mathrm{Y}$ & 5233.3 & $10^\mathrm{e}$  $_{(0.90)}$ & 3 & 19.0\\
\hline
$^{129}\mathrm{Te}$ & 6991.1 & $\frac{3}{2}^\mathrm{e}$  $_{(0.49)}$, $\frac{5}{2}^\mathrm{e}$  $_{(0.49)}$ & 3 & 17.1\\
\hline
$^{151}\mathrm{Sm}$ & 3303.3 & $\frac{33}{2}^\mathrm{o}$  $_{(0.89)}$ & 6 & 21.4\\
\hline
$^{155}\mathrm{Gd}$ & 4087.9 & $\frac{39}{2}^\mathrm{o}$  $_{(0.48)}$, $\frac{41}{2}^\mathrm{o}$  $_{(0.48)}$ & 3 & 17.3\\
\hline
$^{159}\mathrm{Gd}$ & 5943.0 & $\frac{3}{2}^\mathrm{o}$  $_{(0.52)}$, $\frac{1}{2}^\mathrm{o}$  $_{(0.48)}$ & 5 & 14.9\\
\hline
$^{194}\mathrm{Ir}$ &  580.0 & $3^\mathrm{e}$  $_{(0.65)}$, $1^\mathrm{e}$  $_{(0.23)}$, $2^\mathrm{e}$  $_{(0.13)}$ & 5 & 11.1\\
 &  334.1 & $5^\mathrm{o}$  $_{(0.38)}$, $4^\mathrm{o}$  $_{(0.35)}$, $3^\mathrm{o}$  $_{(0.26)}$ & 3 & 16.7\\
\hline
$^{235}\mathrm{U}$ & 1982.4 & $\frac{33}{2}^\mathrm{o}$  $_{(0.29)}$, $\frac{31}{2}^\mathrm{o}$  $_{(0.18)}$, $\frac{29}{2}^\mathrm{o}$  $_{(0.18)}$ & 3 & 18.1\\
\end{tabular}
\end{ruledtabular}
\end{table}
\endgroup

In this work, we proposed a simple data-driven method to find new energy levels from a list of experimentally observed line wavelengths.
Our method is based on the statistical consideration of the energy level candidates computed from the known lower levels and observed wavelengths, where the level candidates from wrong level-line combinations distribute randomly while the candidates from correct combinations may concentrate at the true energy levels.
The use of outlier detection techniques makes the automatic level finding possible.
Based on this method, we found many energy levels of atoms and nuclei from the public database.

One strength of our method is the robustness against a contamination of lines from other species, because such a wrong combination simply falls into the noise floor of the candidate histogram.
By the same reason, emission lines due to other types of transitions, e.g., electric quadrupole or magnetic dipole or even higher multipole transitions, does not alter our method.
The accuracy of the line wavelengths significantly affects the detection power as the noise amplitude linearly depends on the energy resolution.
Thus, by combining a more accurate experiment, our method may lead further level identification of complex systems.

\section*{Appendix: Predictive distribution of wrong candidates}
As can be seen in \fref{fig:tungsten}~(b), the average number of candidates changes over the excited energy $E$ and $J^p$ values of the lower levels.
In order to capture this dependence, we take a moving average of these candidates over an energy interval $\Delta E$ and estimate the value of $\rho$.
For atoms, we use $\Delta E$ = 0.2 eV.
In this interval, we have $K = \Delta E / \delta E$ bins.
Let $k$ be the total number of candidates in this energy interval.
The maximum likelihood estimate of the mean candidate number in each bin is $\rho = k / K$.
However, particularly when $\rho$ is small, this estimate have an uncertainty.

Based on the Bayesian statistics, the posterior distribution of $\rho$ can be described by the Gamma distribution,
\begin{align}
    p_\mathrm{gam}(\rho | k, K) = \frac{K^k}{\Gamma(k)} \rho^{k - 1} e^{-K \rho},
\end{align}
with Gamma function $\Gamma(k) = \int_0^\infty x^{k-1} e^{-x} dx$.
Here, we assume the uniform prior distribution for $\rho$.
The predicted distribution can be obtained by merginalizing $\rho$ as
\begin{align}
    p(r | k, K) = \int_0^\infty p_\mathrm{poi}(r | \rho) p_\mathrm{gam}(\rho | k, K) d\rho,
\end{align}
which gives the negative binomial distribution \eref{eq:nbinom}.

\begin{acknowledgments}
    This work was supported by JSPS KAKENHI Grant Number 19K14680 and 19KK0073.
    K. F. thanks Prof. Masahiro Hasuo for the fruitful discussions.
\end{acknowledgments}

\bibliography{refs}

    
\end{document}